# Astro2020 Science White Paper

# Imaging the Key Stages of Planet Formation

**Thematic Areas:**   ☐ Planetary Systems   ✅ Star and Planet Formation
☐ Formation and Evolution of Compact Objects   ☐ Cosmology and Fundamental Physics
☐ Stars and Stellar Evolution   ☐ Resolved Stellar Populations and their Environments
☐ Galaxy Evolution   ☐ Multi-Messenger Astronomy and Astrophysics

**Principal Author:**
Name:  John Monnier
Institution:  University of Michigan
Email:  monnier@umich.edu
Phone:  734-763-5822

**Co-signers:** (names and institutions)
Gioia Rau   NASA/GSFC
Joel Sanchez-Bermudez   Instituto de Astronomía de la UNAM
Sam Ragland   W.M. Keck Observatory
Rachel Akeson   Caltech/IPAC
Gaspard Duchene   University of California Berkeley
Gerard van Belle   Lowell Observatory
Ryan Norris   Georgia State University
Kathryn Gordon   Agnes Scott College
Denis Defrère   University of Liège
Jacques Kluska   KULeuven
Stephen Ridgway   NOAO
Jean-Baptiste Le Bouquin   University of Grenoble, University of Michigan
Narsireddy Anugu   University of Exeter
Nicholas Scott   NASA Ames
Stephen Kane   University of California, Riverside
Noel D Richardson   University of Toledo
Zsolt Regaly   Konkoly Observatory, Research Center for Astronomy and Earth Sciences, Budapest, Hungary
Zhaohuan Zhu   University of Nevada, Las Vegas
Gautam Vasisht   JPL-Caltech
Keivan G. Stassun   Vanderbilt University
Sean Andrews   Center for Astrophysics | Harvard & Smithsonian
Sylvestre Lacour   Observatoire de Paris
Gerd Weigelt   Max Planck Institute for Radio Astronomy

Neal Turner   Jet Propulsion Laboratory, California Institute of Technology
Fred C Adams   University of Michigan
Douglas Gies   Georgia State University
Nuria Calvet   University of Michigan
Catherine Espaillat   Boston University
Rafael Millan-Gabet   Giant Magellan Telescope Organization
Tyler Gardner   University of Michigan
Chris Packham   University of Texas at San Antonio
Mario Gai   Istituto Nazionale di Astrofisica, Osservatorio Astrofisico di Torino
Quentin Kral   Paris Observatory
Jean-Philippe Berger   IPAG - Université Grenoble Alpes
Hendrik Linz   MPIA Heidelberg
Lucia Klarmann   Max-Planck-Institut für Astronomie
Matthew Bate   University of Exeter, UK
Jaehan Bae   Carnegie Institution of Washington
Rebeca Garcia Lopez   Dublin Institute for Advanced Studies
Antonio Garufi   INAF Arcetri
Fabien Baron   Georgia State University
Mihkel Kama   University of Cambridge
David Wilner   Center for Astrophysics | Harvard & Smithsonian
Lee Hartmann   University of Michigan
Makoto Kishimoto   Kyoto Sangyo University
Johan Olofsson   Instituto de Física y Astronomía, Facultad de Ciencias, Universidad de Valparaíso, Av. Gran Bretaña 1111, Playa Ancha, Valparaíso, Chile
Melissa McClure   University of Amsterdam
Chris Haniff   University of Cambridge, UK
Sebastian Hoenig   University of Southampton (UK)
Michael Line   Arizona State University
Romain G. Petrov   Université de la Côte d'Azur, Nice, France
Michael Smith   University of Kent
Theo ten Brummelaar   CHARA - Georgia State University
Matthew De Furio   University of Michigan
Maria Koutoulaki   Dublin Institute for Advanced Studies
Stephen Rinehart   NASA-GSFC
David Leisawitz   NASA-GSFC
William Danchi   NASA-GSFC
Daniel Huber   University of Hawaii
Ke Zhang   U. Michigan
Benjamin Pope   New York University
Michael Ireland   ANU
Stefan Kraus   University of Exeter
Andrea Isella   Rice U.
Benjamin Setterholm   U. Michigan
Russel White   Georgia State University

**Abstract** (optional):
New images of young stars are revolutionizing our understanding of planet formation. ALMA detects large grains in planet-forming disks with few AU scale resolution and scattered light imaging with extreme adaptive optics systems reveal small grains suspended on the disks' flared surfaces. Tantalizing evidence for young exoplanets is emerging from line observations of CO and H-alpha. In this white paper, we explore how even higher angular resolution can extend our understanding of the key stages of planet formation, to resolve accreting circumplanetary disks themselves, and to watch planets forming **in situ** for the nearest star-forming regions. We focus on infrared observations which are sensitive to thermal emission in the terrestrial planet formation zone and allow access to molecular tracers in warm ro-vibrational states. Successful planet formation theories will not only be able to explain the diverse features seen in disks, but will also be consistent with the rich exoplanet demographics from RV and transit surveys. While we are far from exhausting ground-based techniques, the ultimate combination of high angular resolution and high infrared sensitivity can only be achieved through mid-infrared space interferometry.

INTRODUCTION

In the last 10 years, the study of planet-forming disks has been utterly transformed by high-resolution imaging of gaps, asymmetries, and spirals by mm-wave ALMA interferometry (HL Tau, ALMA Partnership 2015) and adaptive-optics enhanced 8m-class telescopes using coronagraphy (e.g., Garufi et al. 2018). While early SED-based detections of 'transitional disks' opened our imaginations to the possibility that we could study planets as they are forming in real time (e.g., Espaillat et al. 2010), these new detailed images show indeed complex dynamic processes that we still struggle to understand.

Figure 1 shows the results of a hydrodynamic simulation consisting of 4 giant planets that carve out a large gap and would appear as a classic 'transition disk' by its SED (Dong et al. 2015). We see interesting features on the scale of the disk itself (~80 au), large gaps (5-20 au), emission from the inner terrestrial planet formation zone (<4 au), disk gaps and accretion streams caused by individual growing giant planets (<1 au) and finally the circumplanetary accretion disks (<0.2 au) around these planets. To access these scales in nearby star forming region (>100 pc), we need mm-wave and radio interferometry along with diffraction-limited 8m and 30m class telescope imaging. The finest scales (<<1 au) will not be observable by single apertures or ALMA, but will need the resolution afforded by 100-1000m baselines infrared interferometry.

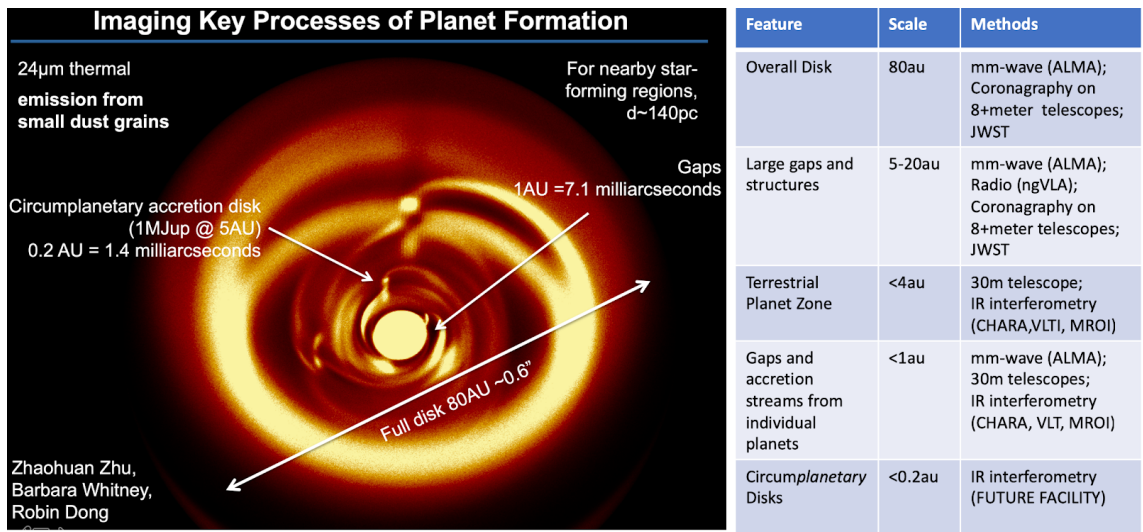

**Figure 1. This radiative transfer model of a complex planet-forming disk shows the key planet-formation processes. ALMA and near-IR imaging with 30m class telescopes can reach down to few AU scales for the closest star forming regions, resolving gaps and possible accreting protoplanets. Near-IR and mid-IR interferometry will be needed to resolve structures within the terrestrial planet forming region and scrutinize the circumplanetary disks themselves.**

Planet formation relies on the interplay of several physical processes involving dust, ice, gas, chemistry, as well as the radiation field from the central star as shadowed by inner disk structures. Observations will be needed to determine the importance of effects such as gravitational instability (Boss et al. 1997), streaming instability (Johansen et al 2007), dust growth (Birnstiel et al. 2010), core accretion (Pollack et al 1996), planetary migration (Tanaka et al 2002), and more. With an accurate theory of planet formation, we hope to explain the observed demographics of exoplanets as pieced together from RV, transit, and direct-imaging surveys.

CURRENT STATUS and ISSUES

Figures 2 & 3 show the powerful datasets now available to modellers. The combination of ALMA, large telescope adaptive optics, and long baseline IR interferometry allows an unprecedented view of planet formation, letting astronomers peek in on young disks that are actively forming planets. These data are now sufficiently comprehensive to permit detailed 3D radiative transfer coupled with hydrodynamical simulations. The source in Figure 2 (HD 163296) is especially interesting since giant exoplanets might exist here, based on signatures in ALMA CO data (Teague et al. 2018). It is worth noticing that this object was a generic young star based on its spectral energy distribution (SED) and not identified as "transitional" -- it seems that probably all young stellar objects harbor "transitional disks" of some kind, with active and ongoing planet formation. The potential for advancing astronomy is immense even just by exploiting our current capabilities, let alone with an upgraded ALMA, ngVLA, 30m class ELTs, and kilometric-baseline IR interferometers being studied. Figure 3 is the tip of the iceberg.

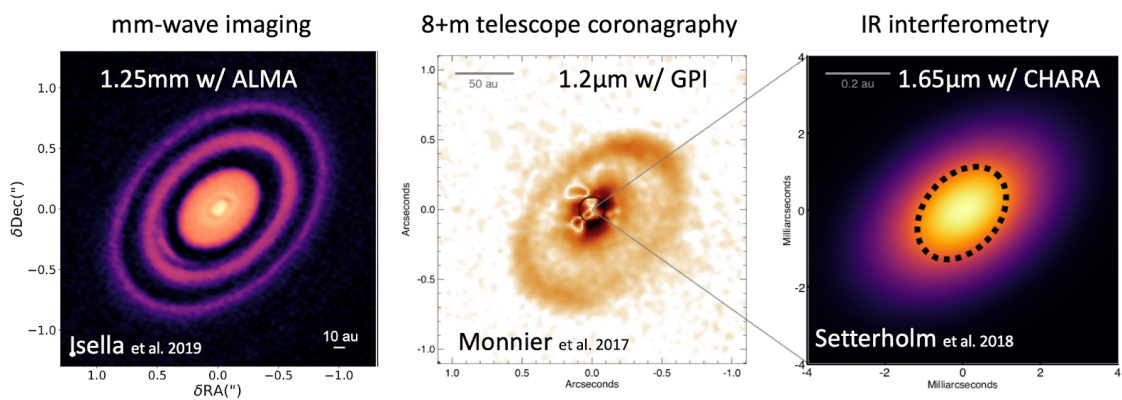

**Figure 2. Multi-wavelength and multi-scale study of HD 163296 (MWC 275). (left) Recent mm-wave imaging from ALMA shows multiple rings and some unusual structure (vortex?) in the middle ring. (middle) On the same scales as ALMA, the Gemini Planet Imager mapped out scattering on the disk surface. (right) CHARA interferometer measured the innermost hot dust within a 1 au, a region that we know relatively little about but which can affect the outer disk through shadowing. In the next decade, we will be collecting this kind of powerful dataset for hundreds of YSOs in the nearest star forming regions.**

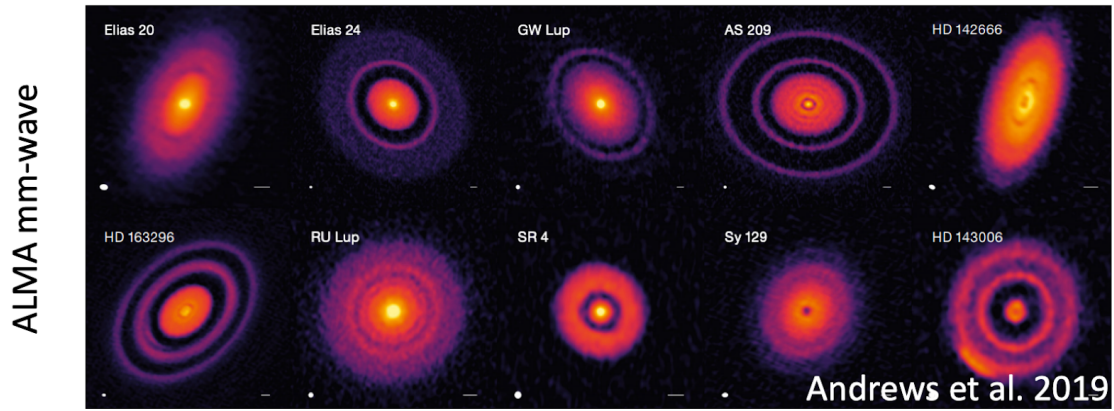
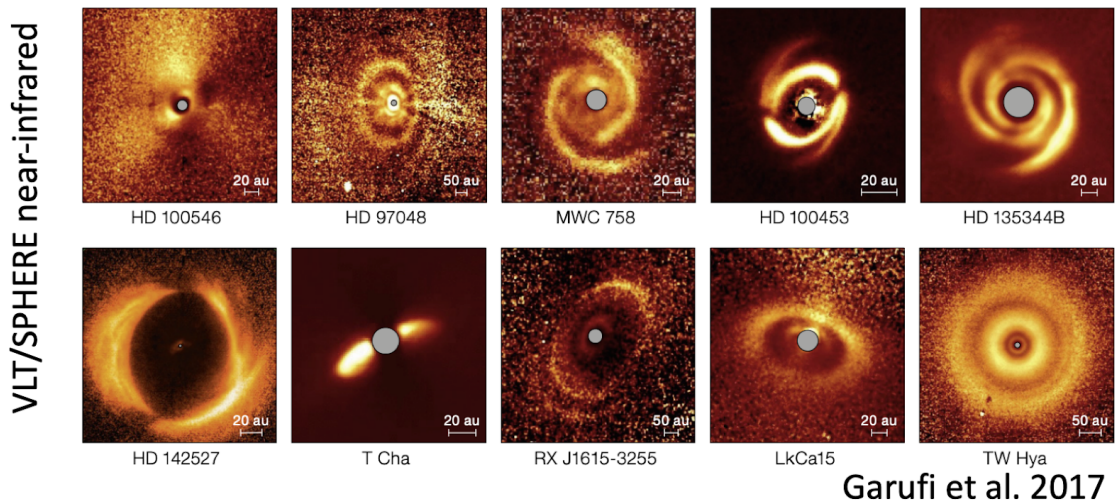

Figure 3. Surveys of planet-forming disks by ALMA, VLT/SPHERE, and others instruments are just starting. The diversity of structures are incredible and we are seeing rapid evolution in our understanding of planet formation theory.  We will continue to see huge advances as an upgraded ALMA, ngVLA, new 30m-class Extremely Large Telescopes (ELTs), JWST,  and more sensitive IR interferometers come online in the next decade.

FUTURE ADVANCES

While the growing galleries of ALMA and NIR images are incredible, we are still relatively blind in the mid-infrared (MIR, 3-20 microns) and in the inner few AU of the disk.  These warm inner regions are where the terrestrial planets form, thus direct imaging here is important for understanding rocky planet formation. Unfortunately, the MIR suffers from relatively poor angular resolution due to the wavelength-dependence of the diffraction-limit ( $\Theta = \lambda/D$ ) and high thermal backgrounds from sky emission.  While ground-based telescopes/interferometers have high angular resolution (0.05-0.25"), they also have high thermal backgrounds that significantly reduce sensitivity; conversely, cooled space telescopes (e.g., Spitzer, WISE) have very low backgrounds but their smaller diameters deliver poor angular resolution.

The Planet Formation Imager Project has been exploring the science potential of **milliarcsecond imaging** in the mid-infrared through the help of simulations. Since such fine angular resolution requires >100m size telescopes, we currently consider infrared interferometry only. Figure 4 shows that key stages in Jupiter formation for a solar system analogue disk could be directly imaged with a next-generation MIR-optimized interferometer. By extending the capabilities of ALMA to 100x shorter wavelengths, we can continue the dramatic progress in understanding how our solar system was formed and how variations in planet formation conditions might yield the diverse demographics of exoplanets measured by RV and transit surveys.

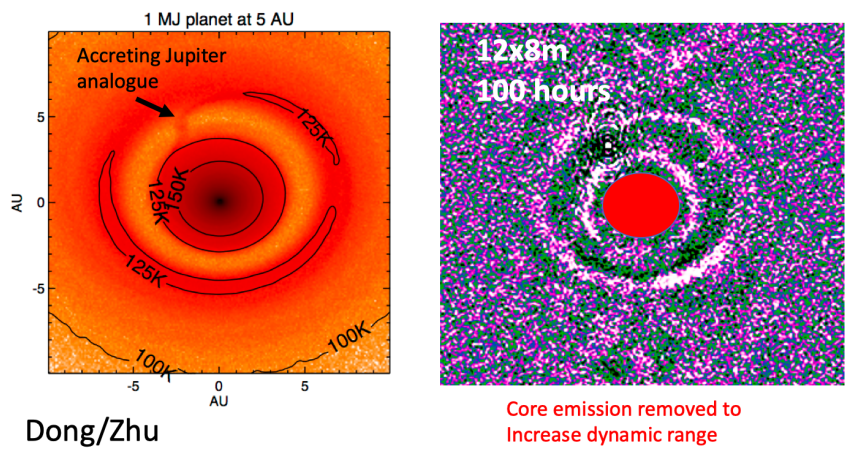

**Figure 4.** (left) This radiative-transfer synthetic image (wavelength 10 microns) shows the inner 20x20au of a solar analogue protoplanetary disk, based on a realistic hydrodynamic simulation of 1 $M_J$ planet forming at 5 au. We see an au-scale gap and the circumplanetary accretion disk itself. For a nearby star forming region (140 pc), this entire region would fit in about 2x2 pixels even for the 39m E-ELT. We wanted to image this region using a simulation of an ambitious ground-based interferometer of 12x 8m class telescopes with 1 km baselines. (right) This is the resulting image reconstruction for a simulated 100-hour observing sequence. We clearly resolve the planet-induced gap with <1 au angular resolution and detect the mid-IR circumplanetary disk emission (from PFI Planet Formation Imager study, presented by Monnier et al. 2018).

Since we are boldly looking into the future, we can go further in angular resolution. If one can obtain <1 milliarcsecond resolution in L band (3.8 microns), **we could start to resolve the circumplanetary disks of forming giant planets!** Not only is this exciting for understanding the details of how giant planets (and their moons) form, we could use line emission to measure Keplerian motion of disk gas -- this would be a monumental breakthrough that would allow us to measure young exoplanet masses.

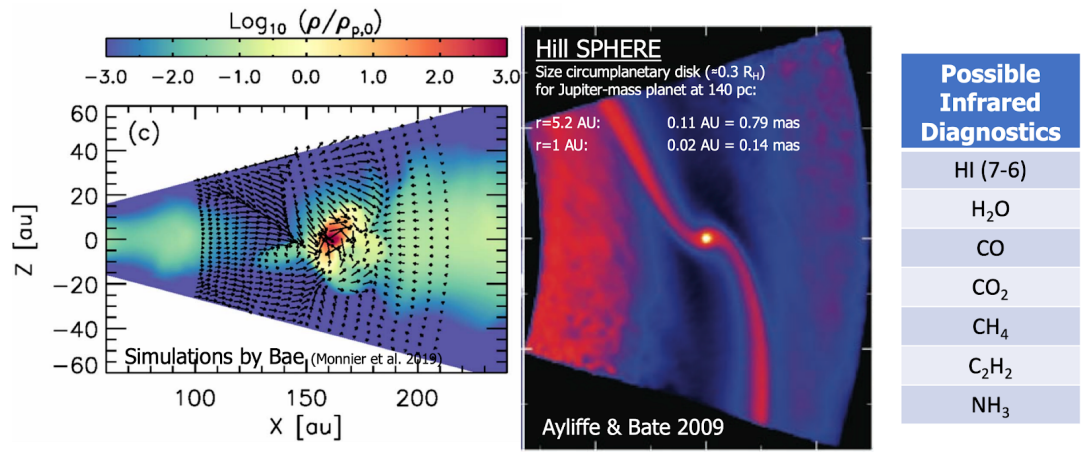

Figure 5. (left) Hydrodynamical simulation by Bae (found in Monnier et al. 2019) of a forming massive exoplanet. (middle) Planet-forming simulation by Ayliffe & Bate (2009) showing gap-formation as a giant planet forms in a young disk. The circumplanetary accretion disk has the characteristic size of ⅓ the Hill Sphere, corresponding to <1 milliarcsecond for Jovian analogues in the nearest star-forming regions. (right) List of abundant species with strong spectral features in the 3-15 micron wavelength range. These lines could be used to track the kinematics of circumplanetary disk gas, a chance for a direct exoplanet mass measurement. While end-to-end simulations are still lacking, space-based interferometry will likely be needed to attain the needed signal-to-noise ratio within spectral lines for mass measurements.

Figure 5 shows some of the stages of giant planet assembly and some example of recent hydrodynamic simulations. This aspect of giant planet formation is extremely poorly understood and current simulations are highly preliminary. Nonetheless, astronomers have started to identify some key tracers for warm gas in the vicinity of an accreting Jovian planet. The next decade will see a dramatic computational improvement in this area, with solid links developed between astronomical observations of actively accreting giant exoplanets and properties of the our solar system giant planets and their moon systems.

SUMMARY
Diffraction-limited imaging using mm-wave/IR interferometry and large ground-based telescopes are revolutionizing our understanding of how planets form. We are in the middle of this fast-paced, paradigm-shifting period and it will take a decade to fully take advantage of current facilities. New facilities, such as an upgraded ALMA, ngVLA, and the 30m-class ELTs, promise even more exciting breakthroughs and we highlight that current technologies could support a new mid-infrared interferometer capable of imaging the terrestrial planet forming region with sub-au resolution. Ultimately we need the combination of low thermal background and long baselines to resolve the individual **circumplanetary** disks, and we recommend forward-thinking new investments into space interferometry technology this decade to pave the way.